# Observation of Optical Refrigeration in a Holmium-doped Crystal

Saeid Rostami, Alexander R Albrecht, Azzurra Volpi, and Mansoor Sheik-Bahae[*]

*University of New Mexico, Department of Physics and Astronomy, 1919 Lomas Blvd. NE, Albuquerque, NM 87131, USA*

[*] **Corresponding author**

Correspondence to Mansoor Sheik-Bahae; e-mail: msb@unm.edu

**ABSTRACT**
We report the first demonstration of solid-state optical refrigeration of a Ho-doped material. A 1 mol% Ho-doped Yttrium Lithium Fluoride (YLF) crystal is cooled by mid-IR laser radiation, and its external quantum efficiency and parasitic background absorption are evaluated. Using detailed temperature-dependent spectroscopic analysis the minimum achievable temperature of a 1% Ho:YLF sample is estimated. Owing to its narrower ground- and excited-state manifolds, larger absorption cross-section, and coincidence of optimum cooling wavelength of 2070 nm with commercially available high-power and highly efficient Tm-fiber lasers, $Ho^{3+}$-doped crystals are superior to $Tm^{3+}$-doped systems for mid-IR optical refrigeration. With further improvement in material purity and increased doping concentration, they offer great potential towards enhancing the cooling efficiency nearly two-fold over the best current Yb:YLF systems, achieving lower temperatures as well as for realization of eye-safe mid-IR high-power radiation balanced lasers.

**INTRODUCTION**

Optical refrigeration relies on anti-Stokes fluorescence[1] where absorption of low-entropy and coherent light (i.e. laser) with photon energy $h\nu$ is followed by efficient emission of high-entropy spontaneous emission (fluorescence) with a mean energy of $h\nu_f > h\nu$ as shown schematically in Fig. 1a. This excess energy in fluorescence must be extracted from the internal energy of the system (e.g. phonons), thus leading to net cooling. The optical cooling efficiency $\eta_c$, defined as the heat-lift per absorbed photon, is expressed as[2]:

$$\eta_c = p \frac{h\nu_f}{h\nu} - 1, \qquad (1)$$

where $p \lesssim 1$ denotes the probability that an absorbed pump photon will lead to a fluorescence photon *exiting* the system. Practical consideration on the pump detuning requires $h\nu_f - h\nu \sim O(kT)$ which in turn entails $p > 98\%$ assuming ~1 eV transition at room temperature. In 1950, Kastler[3] proposed rare-earth (RE) doped solids as a suitable candidate for fluorescence cooling due to their extremely high fluorescence efficiency. The first observation of optical refrigeration, however, did not materialize until 1995, when researchers at Los Alamos National Laboratory reported net cooling of a high-purity Yb-doped ZBLANP glass[4] at a laser wavelength $\lambda$=1030 nm. Since then, considerable advances[5–8] have been made towards achieving cryogenic operation by exploiting high-purity crystalline host materials such as $YLiF_4$ (YLF)[9,10]. Most recently, a Yb:YLF crystal was cooled to below 90 K from room temperature[11]. Another major milestone in optical refrigeration reported this year involved the long-awaited tangible application of this technology: an *arbitrary* load (in this case a HgCdTe IR sensor) was cooled to 135 K which marks the first realization of an all-solid-state cryogenic refrigerator device with a universal cold-finger[12]. Such cryocoolers offer certain advantages over their existing mechanical counterparts, as they are virtually vibration-free and can have superior reliability and lifetime, due to their lack of moving parts. Currently, a drawback of this technology is its rather low efficiency (<1%). This limitation, however, is not a fundamental one: it has been suggested that with larger detuning (in extremely high purity crystals) and/or fluorescence harvesting (e.g. using photovoltaic convertors[13] or Stokes shifters[14]), this is only limited by the Carnot efficiency[14]. An obvious alternative to enhancing the efficiency is to use lower energy transitions. This becomes apparent by rewriting Eq. 1 assuming $p \sim 1$ and a detuning of $\sim kT$, which gives $\eta_c \approx kT/h\nu$, signifying an inverse scaling of cooling efficiency with photon energy. This scaling law was first validated by cooling 1% Tm:ZBLANP glass at $\lambda \sim 2000$ nm, thus improving the efficiency nearly two-fold over Yb-doped systems[15]. Later, $Tm^{3+}$:$BaY_2F_8$ (Tm:BYF) crystals were also cooled successfully with improved performance[16].

Another advantage of mid-IR optical refrigeration is material purity. Based on Yb cooling results[17], the current assumption is that the major contamination is caused by transition metals (e.g. iron[18]), which have much smaller absorption cross-sections in the mid-IR compared to the near-IR[19].

A figure of merit for any cooling-grade material is its so-called minimum achievable temperature (MAT). This quantity is defined by the lowest temperature at which $\eta_c$ vanishes and subsequently turns negative (i.e. enters the heating regime). MAT is not a fundamental property of a dopant or host, rather it is highly sensitive to the growth quality and purity of any given material. This further becomes apparent by examining the constituents of $\eta_c$, namely $\nu_f$ and $p$, and their variation with temperature. It is straight forward to show that deviation of $p$ from unity is due to the ubiquitous presence of both nonradiative decay and unwanted impurities that cause parasitic heating. That is, $p(\nu,T) = \eta_{ext}\eta_{abs}(\nu,T)$, where $\eta_{ext} = (1 + W_{nr}/\eta_e W_r)^{-1}$ is the external quantum efficiency (EQE) and $\eta_{abs}(\nu,T) = (1 + \alpha_b/\alpha_r(\nu,T))^{-1}$ denotes the absorption efficiency[2,14]. Here $W_r$, $W_{nr}$, $\alpha_b$, and $\alpha_r(\nu,T)$ are radiative and nonradiative decay rates, background, and resonant absorption coefficients, respectively. Additionally, the radiative (spontaneous emission) rate is effectively suppressed by the fluorescence escape efficiency $\eta_e$, which takes into account the effect of fluorescence reabsorption as well as radiation trapping via total internal reflection. It is quite reasonable to assume that $\eta_{ext}$ is only weakly temperature dependent. The overriding temperature dependence of $p(\nu,T)$ arises from $\alpha_r(\nu,T)$ for wavelengths near the optimum cooling efficiency, corresponding to transitions originating from the top of the ground-state. Under the plausible assumption that Boltzmann quasi-equilibrium establishes in each manifold prior to spontaneous emission, this temperature dependence follows $\alpha_r(\nu,T) \propto \left(1 + e^{\delta E_{gs}/kT}\right)^{-1}$, where $\delta E_{gs}$ is the width of the ground-state manifold[14].

Similarly, with Boltzmann quasi-equilibrium established in the excited-state manifold, the mean fluorescence energy redshifts as the temperature of the crystal is lowered according to $h\nu_f(T) \sim h\nu_f(0) + \delta E_{es}/(1 + e^{\delta E_{es}/kT})$, where $\delta E_{es}$ denotes the width of the excited-state manifold[14]. Such functional dependences on the widths of energy manifolds are indeed the key factors that have rendered YLF a suitable host material for optical refrigeration due to its rather weak Stark crystal field acting on RE dopant ions. Similarly, the main motivation of investigating $Ho^{3+}$ is the fact that it has narrower ground and excited state manifolds ($\delta E_{gs}$=303 cm$^{-1}$, $\delta E_{es}$=140

cm$^{-1}$) compared to Tm$^{3+}$ ($\delta E_{gs}$=419 cm$^{-1}$, $\delta E_{es}$=373 cm$^{-1}$) for a given host[20]. It is also worth noting that the manifolds in Ho$^{3+}$ are narrower than those in Yb$^{3+}$ (YLF)[21] with $\delta E_{gs}$=449 cm$^{-1}$, $\delta E_{es}$=278 cm$^{-1}$, providing further advantage of Ho for cryogenic refrigeration. In the following we describe experimental details of laser cooling of a 1% Ho:YLF crystal; this represents the first observation of laser cooling of any Ho-doped solid. We will discuss the prospects of a Ho-doped crystal for cryogenic cooling and the necessary conditions to obtain higher cooling efficiency than in Yb-based systems, as well as its potential for mid-IR radiation balanced lasers.

## RESULTS
### Room-Temperature Analysis

We used a 4.8×4.8×5.0 mm$^3$ high purity 1% Ho$^{3+}$-doped YLF crystal grown by Czochralski process (AC Materials, Tarpon Springs, FL). The measured room temperature absorption coefficient and emission spectra associated with the transitions between the ground-state ($^5I_8$) and the first excited state ($^5I_7$) Stark manifolds are shown in Fig. 1b. The shaded area in the absorption spectrum is the so-called "cooling tail" with $\lambda > \lambda_f$, where $\lambda_f = c/\nu_f \sim 2015$ nm denotes the mean fluorescence wavelength. To quantify the sample's EQE and $\alpha_b$, and to investigate if the sample is of "cooling grade", a mid-IR laser source of modest power (1-2 W), narrow linewidth (<1 nm), and tunable in the vicinity of $\lambda_f$ is required. For this purpose, we designed and constructed a singly-resonant continuous-wave (CW) optical parametric oscillator (OPO) based on temperature-tuned periodically-polled lithium niobate (PPLN)[22,23]. This OPO, pumped by a high-power CW fiber laser at 1070 nm, can be tuned from 1900 nm to 2300 nm, and is further detailed in the Materials and Methods section.

The cooling efficiency $\eta_c$ of the Ho:YLF sample is evaluated at room temperature by measuring the temperature change ($\Delta T$) induced by irradiating the sample with the OPO output, as the wavelength is tuned from below to above $\lambda_f$. Under steady-state condition and for small temperature changes, $\Delta T(\lambda) = K\eta_c(\lambda)P_{abs}(\lambda)$, where $P_{abs}$ is the absorbed laser (OPO) power, and $K$ is a constant (scaling factor) that varies inversely with the thermal load on the sample. Therefore, $\eta_c(\lambda) = \Delta T/KP_{abs}$ can be extracted by measuring $\Delta T$ and the absorbed power as function of $\lambda$. The aforementioned method is termed Laser-Induced Thermal Modulation Spectroscopy (LITMoS)[24,25]. Figure 2a shows the schematic of a typical LITMoS experiment in which the cooling sample is positioned on top of two transparent low thermal conductivity holders inside a vacuum chamber (10$^{-6}$ torr) in order to minimize the conductive and convective thermal loads and thus maximize $\Delta T$ induced by a double pass of the OPO beam through the sample. This is particularly useful at enhancing the signal-to-noise ratio in the long wavelength regime, where the absorption coefficient (and hence $P_{abs}$) drops drastically. The relative temperature change $\Delta T$ of the sample is measured using an IR thermal camera (Thermal Eye Nanocore 640 L3-Communications Corporation, TX, USA) which views the sample from outside the vacuum chamber through a KCl window.

The temperature change of the crystal at each wavelength is extracted from thermal images following standard image processing that involves spatial and temporal averaging. Care is taken to ensure that thermal camera response is linear, and that the ambient temperature remains constant during the experiment. The absorbed power is calculated from the transmitted pump power and the room-temperature absorption coefficient, taking Fresnel reflections into account. The normalized data $\Delta T(\lambda)/KP_{abs}(\lambda)$ is then fitted with Eq. 1 by adjusting K, $\eta_{ext}$, and $\alpha_b$. Note that, for simplicity, we assume $\alpha_b$ does not vary with wavelength within the narrow (~200 nm) spectral range of interest. This assumption is adopted primarily since the origin of the parasitic absorption is not precisely known from sample to sample as, for example, it could arise from a variety of transition metals or other rare-earth ions[18]. It is further assumed that $\alpha_b$ is also temperature independent. The validity of both of these assumptions and their implication in cryocooling experiments will be revisited later in this letter.

The measured LITMoS test on the 1% Ho:YLF sample at room temperature ($T$=300 K) along with its corresponding fitting parameters $\alpha_b$, $\eta_{ext}$ are shown in Fig. 2b. A net-cooling window is observed between $\lambda_{c1}$=2059 nm and $\lambda_{c2}$=2215 nm. Generally, in high purity samples where $\alpha_b<10^{-3}$ cm$^{-1}$, $\eta_{ext}$ can be estimated with a fair degree of accuracy from $\lambda_f/\lambda_{c1}$, while the value of $\alpha_b$ is highly sensitive to the location of $\lambda_{c2}$. The best fit to data, as shown in Fig. 2b, gives $\eta_{ext}$=98.0±0.2% and $\alpha_b$=(5±2)×10$^{-5}$ cm$^{-1}$ for this 1% Ho:YLF sample.

**Low-Temperature Analysis**

Thus far, we have demonstrated that this Ho:YLF sample is of cooling-grade at room temperature, exhibiting maximum cooling at a wavelength of $\lambda$ =2150 nm. The next task at hand is to identify its potential for cryocooling operation by evaluating its MAT. This in turn necessitates evaluation of $\eta_c(\lambda,T)$ down to cryogenic temperatures. As stated earlier, starting with the assumptions that $\eta_{ext}$ and $\alpha_b$ are temperature independent, we only need to obtain the temperature dependence of the remaining ingredients of $\eta_c$, namely $\alpha_r(\lambda,T)$ and $\lambda_f(T)$. The latter is measured by recording the fluorescence spectra $S(\lambda)$ associated with the first excited-state transition as the temperature of the sample, placed in a cryostat, is varied from 300 K to 80 K. The emission spectra are collected by a fiber-coupled mid-IR optical spectrum analyzer (Thorlabs OSA203B). Since YLF is an anisotropic (uniaxial) crystal, the spectra for both $\pi$ (E∥c) and $\sigma$ (E⊥c) polarizations are separately recorded at each temperature by using appropriate polarizers in front of the collection fiber. The mean fluorescence wavelength $\lambda_f^{\pi,\sigma}(T) = \int \lambda S^{\pi,\sigma}(\lambda,T)d\lambda/\int S^{\pi,\sigma}(\lambda,T)d\lambda$ is calculated for each polarization followed by evaluating the exiting total mean fluorescence wavelength by performing a weighted average along the three Cartesian axis of the crystal - given by $\lambda_f(T) = (2/3)\lambda_f^\sigma(T) + (1/3)\lambda_f^\pi(T)$. The remaining task is now to determine the temperature-dependent resonant absorption coefficients $\alpha_r(\lambda,T)$ by utilizing the reciprocity theorem[26–28] and the McCumber relation which gives $\alpha_r(\lambda,T) \propto \lambda^5 S(\lambda,T)e^{hc/\lambda kT}$. Proportional spectra thus calculated are then calibrated to a directly measured absorption value (e.g. at $\lambda$=2055 nm) to get the exact absorption spectra $\alpha_r(\lambda,T)$ at each temperature. The absorption spectra that are obtained using reciprocity agree well with those directly measured using an FTIR, while having the advantage of exhibiting less noise in the long wavelength ($\lambda$>2100 nm) tail, which is of particular interest for laser cooling.

Figure 3a shows the measured temperature variation of $\lambda_f$ for the 1% Ho:YLF crystal. The data is normalized to its room temperature value $\lambda_f^{Ho}$(300 K) = 2015 nm for comparison with the same measured quantity for a cooling grade 1% Tm:YLF crystal[22,24] having $\lambda_f^{Tm}$(300 K)=1822 nm. A redshift of ~1% is seen in Ho:YLF compared to ~3.5% in Tm:YLF as the temperature is varied from 300 K to 80 K. As stated earlier, this signifies the narrower width of the excited-state manifold in Ho which makes it highly suitable for cryogenic cooling. Additionally, a nearly 2-fold enhancement in the absorption cross section of Ho$^{3+}$ over Tm$^{3+}$ makes the case of Ho-based crycoolers in mid-IR even stronger[20,27]. The calibrated absorption spectra for E∥c obtained from the reciprocity relation in the same temperature range are given in Fig. 3b. This polarization was chosen, as Ho:YLF exhibits larger absorption cross-section for E∥c at wavelengths in the cooling tail.

With $\eta_{ext}$, $\alpha_b$, $\lambda_f(T)$, and $\alpha_r(\lambda,T)$ known, we plot $\eta_c(\lambda,T)$ to identify the cooling and heating spectral regimes at all temperatures, and subsequently obtain the value of MAT for this crystal. Fig. 3c shows the map of $\eta_c$ versus $T$ and $\lambda$. The white demarcation in this plot signifies $\eta_c$=0 and marks the minimum achievable temperature (MAT) for the corresponding excitation wavelength. The lowest (or global) MAT for this crystal is ~130±10 K at $\lambda$~2070 nm which corresponds to the $E_{12} \rightarrow E_{13}$ transition between $^5I_8$ and $^5I_7$ manifolds (Fig. 1a) in $Ho^{3+}$-doped YLF. It is also worth noting that this wavelength conveniently coincides with readily available high power Tm-and Ho-fiber lasers[29,30]. Additionally, Tm-fiber (or disk) lasers, when pumped by high power diodes near 790 nm, are known to be highly efficient (~65%) due to the well-known 2-for-1 cross-relaxation pumping scheme[31,32].

While future efforts must focus on lowering the MAT beyond 130 K, we should recall that this value of MAT was estimated following the assumption that $\alpha_b$ was independent of temperature. However, recent experiments in Yb:YLF crystals have revealed that this assumption must be revisited; these crystals have been cooled to lower temperatures than predicted by the constant $\alpha_b$ models[11,17]. In particular, $\alpha_b$ in a 5% doped Yb:YLF sample was shown to reduce by nearly an order of magnitude as the temperature is lowered from 300 K to 100 K, which in turn lowered MAT from about 110 K to below 90 K, in excellent agreement with experimental results[11,17]. Such temperature dependence in $\alpha_b$ cannot be generalized, since the parasitic absorption can arise from a variety of contaminants; however, it is not unreasonable to reassess the MAT in Ho:YLF assuming a similar dependence. For example, lowering $\alpha_b$ to $1\times10^{-5}$ cm$^{-1}$ further reduces the MAT to about 100 K for the current Ho-doped sample. As the absorption efficiency depends on $\alpha_b/\alpha_r$, even further improvement in MAT can be achieved by increasing the doping concentration. Investigations on excited state dynamics of $Ho^{3+}$ ions in ZBLAN glass[33] show that concentration quenching of the radiative decay from the $^5I_7$ manifold only sets in at ~4% doping.

### DISCUSSION

We have demonstrated optical refrigeration in a Ho-doped material in the mid-IR for the first time. This offers multiple potential advantages over existing Yb-doped systems for cryogenic cooling as well as for realization of high-power mid-IR RBLs.

Optical refrigeration, since its first demonstration, has been touted as a mechanism for realizing all-solid-state cryocoolers without any moving parts and vibrations. Such a device was recently demonstrated by cooling an IR sensor to <135 K by using Yb:YLF as the cooling element[12]. An essential requirement for a practical optical cooler is that the load or cold-finger must be efficiently shielded from the intense fluorescence emanating from the cooling crystal using a delicately designed thermal link[12]. The next generation of Yb-based cryocoolers are to be integrated with the NIST single-crystal Si reference cavities that need to be cooled to 124 K in a totally vibration-free environment[34,35]. A mid-IR based optical refrigerator can be highly beneficial for this application, since the load (Si cavity) is transparent to the ~2 μm fluorescence, and therefore the thermal link can be eliminated altogether.

As described by Eq. 1 and the ensuing discussion, mid-IR optical refrigeration can potentially offer enhanced cooling efficiency due to energy scaling and other characteristic advantages of the system. The current material with 1% doping and $\eta_{ext}$~98% does not yet match the cooling efficiency of our best Yb:YLF system with 10% doping concentration and $\eta_{ext}$>99%. However, Ho-doped crystals having narrower ground-state manifold, higher absorption cross section, and lower parasitic background absorption promise to outperform Yb:YLF. A modest improvement in $\eta_{ext}$ would allow the $E_{12} \rightarrow E_{13}$ resonance in Ho:YLF occurring at ~2065 nm to be accessed, thus leading to considerable cooling efficiency enhancement. This can be achieved through high purity growth of Ho:YLF crystals or using other host materials with lower phonon energies such as BaY$_2$F$_8$ (BYF) which could further suppress the multi-phonon relaxation mechanism. Recent studies on Tm-doped crystals show that BYF[23,24] with a phonon energy of 350 cm$^{-1}$ (i.e. 100 cm$^{-1}$ lower than in YLF)[36], improves $\eta_{ext}$ from 0.980 to >0.995. Moreover, a reasonable and modest increase in the dopant concentration[33] would further enhance the cooling efficiency above that of Yb-doped systems. To illustrate this, we have evaluated the maximum cooling efficiency of Ho-doped crystal $\eta_{c(max)}^{Ho}$-subject to such minor modifications- relative to the efficiency $\eta_{c(max)}^{Yb}$ for the best Yb-doped system (10% Yb:YLF, $\eta_{ext}$=0.996)[10]. The ratio of these cooling efficiencies versus crystal temperature down to 100 K are shown in Fig. 3d. We have kept the background absorption coefficient $\alpha_b$ in both crystals the same as their measured room temperature values. We note that if $\eta_{ext}$ in 1% Ho:YLF (or BYF) were to improve from 0.98 to 0.99, it would already match the performance of the 10% Yb:YLF at low temperatures of interest (e.g. 150 K). Combined with the availability of higher efficiency mid-IR lasers at the optimum cooling wavelength (e.g. Tm-fiber lasers), this promises an enhanced wall-plug efficiency. Moreover, increasing the doping concentration to 2% can lead to a 2-fold efficiency enhancement of Ho- over Yb-doped systems at 150 K (i.e. 0.24% over 0.13%)[10]. The enhancement in the cooling efficiency ratio seen at lower temperatures is a consequence of Yb:YLF approaching its MAT ($\eta_c$= 0) at a higher temperature than Ho assuming improved doping and quantum efficiency. As discussed earlier, recent experiments suggest a strong reduction of the background absorption at low temperatures which in turn will lead to lower MATs for these materials[17]. In short, the confluence of lower energy gap, narrower ground- and excited state manifolds, higher absorption cross section, and lower parasitic absorption makes Ho-doped crystals potentially superior to Yb-doped systems for cryogenic optical refrigeration with reasonable improvements in material synthesis.

A different yet promising application of anti-Stokes fluorescence cooling is more concerned with the removal of heat rather than cooling to low temperatures. This process is particularly attractive in lasers where the laser action and cooling correspond to the same atomic transition[37]. The principle of such "athermal" lasers was proposed by Bowman[38] and was termed "radiation balanced lasers" (RBL). RBL operation was soon demonstrated in Yb:YAG rods with CW powers approaching 500 W and free from any thermal distortion at 1050 nm[39]. Identifying high quality cooling grade materials in mid-IR, as reported in this letter, paves the way for realization of high power eye-safe RBLs at 2 μm[23,24]. A mid-IR RBL may be realized by pumping at $\lambda_P$=2070 nm with laser operation at $\lambda_L$~2090 nm. Improving $\eta_{ext}$ and the doping concentration not only enhances the cooling efficiency for refrigeration applications, as previously discussed, but equally improves the optical-to-optical efficiency and maximum power density that can be extracted from such a laser under radiation-balanced operation.

In summary, we demonstrated net optical refrigeration in a Ho-doped material for the first time. External quantum efficiency ($\eta_{ext}$=98%), parasitic background absorption ($\alpha_b$=5×10$^{-5}$ cm$^{-1}$), and temperature-dependent emission and absorption spectra were carefully measured for a 1% Ho:YLF crystal. Subsequently, the cooling efficiency as function of wavelength and temperature was evaluated which in turn led to an estimation of the minimum achievable temperature of 130±10 K at $\lambda$~2070 nm. We conclude that Ho:YLF is superior to Tm-doped crystals for mid-IR cryocooler applications due to narrower energy manifolds, availability of high-power lasers, and larger absorption cross-section at the optimum cooling wavelength. Furthermore, we showed that with these spectral characteristics Ho-doped crystals may outperform Yb-doped systems, assuming reasonable improvements in doping and quantum efficiency. Finally, the utility of Ho for high-power, mid-IR, eye-safe radiation-balanced lasers was discussed.

## MATERIALS AND METHODS

A mid-IR continuous-wave (CW) optical parametric oscillator (OPO) was designed and constructed to serve as the tunable pump source for optical refrigeration in Ho:YLF. The nonlinear crystal was a L=50 mm long temperature-tuned MgO-doped periodically poled lithium niobate (MgO:PPLN) crystal with multiple gratings that provide quasi phase-matching (QPM). The pump laser was a CW Yb:fiber laser at 1070 nm (IPG, YLR-500-SM). The OPO cavity is a singly-resonant standing-wave V-cavity (Fig. 4a) formed by two concave mirrors (M1 and M2) having radius of curvature ROC=-20 cm, and a flat output coupler (OC). The OPO operates in singly-resonant mode with M1 and M2 having broadband high reflectance in the signal range (1.7-2.1 μm) and high transmission for both pump (1070 nm) and the idler (2.2-2.8 μm). The signal was coupled out from OC with ~4% transmission. Since our pump laser at 1070 nm has a relatively broad linewidth ($\gtrsim$ 1 nm), resonating the signal (instead of the idler), allows improving the pump acceptance bandwidth to 1 nm[40]. The pump laser is focused to a waist size of $w_0$~70 μm at the middle of the PPLN; this corresponds to $L/2z_0$ ~2.3 where $z_0$ is the Rayleigh range. A pair of intracavity etalons (300 μm thick UV fused silica and 500 μm thick ZnSe) were inserted to reduce the linewidth of the OPO to <0.5 nm. The tuning curve of the OPO versus PPLN temperature is shown in Fig. 4b for a grating period Λ=31.9 μm. Fig. 4c shows the corresponding normalized spectra of the narrow linewidth signal and idler indicating the tuning range used for optical refrigeration of the 1% Ho:YLF crystal.

## ACKNOWLEDGMENTS
The authors acknowledge support from Air Force Office of Scientific Research (FA9550-15-1-0241 and FA9550-16-1-0362 (MURI)) and U.S. Army "Efficient Materials for Optical Cryocoolers" (W911SR-17-C-0039). We thank Dr. Richard Epstein (ThermoDynamic Films LLC), Dr. Zhou Yang (UNM) and Dr. Brian Walsh (NASA) for useful discussions.

## CONFLICT OF INTERESTS
The authors declare no conflict of interests.

## CONTRIBUTIONS
M.S-B., S.R. and A.R.A. designed the experiments. S.R. performed the experiments with assistance of A.V., wrote the main manuscript text and prepared the figures. M.S-B., A.V., and A.R.A supervised the efforts. All authors contributed to the analysis and reviewed the manuscript.

**FIGURE LEGENDS**

**Figure 1: Optical refrigeration process.** (a) Anti-Stokes fluorescence cooling process in $Ho^{3+}$ ions, (b) emission (red line) and absorption (blue line) spectrum of 1% Ho:YLF crystal at $T$=300 K ($\lambda = c/\nu$). The shaded region denotes the cooling tail ($\lambda > \lambda_f$=2015 nm). Emission spectrum is measured with a scanning optical spectrum analyzer under laser excitation at 1890 nm. The absorption spectrum is directly measured with an FTIR spectrometer under E∥c configuration (c is the optical axis).

**Figure 2: Experimental setup and LITMoS data.** (a) Schematic of mid-IR laser cooling and LITMoS test setup for Ho-doped crystals, (b) LITMoS test result for 1% Ho:YLF crystal; the theoretical fit to the data, using Eq. 1, gives the external quantum efficiency ($\eta_{ext}$) and the parasitic (background) absorption coefficient ($\alpha_b$). The insets show two thermal images corresponding to heating and cooling regimes.

**Figure 3: Minimum Achievable Temperature.** (a) Temperature dependence of the mean fluorescence wavelength ($\lambda_f$) for cooling grade 1% Ho:YLF and 1% Tm:YLF crystals. For comparison, data are normalized to room temperature values. (b) Temperature dependence of the resonant absorption coefficient of $^5I_8$-$^5I_7$ transition in 1% Ho:YLF from 300 K to 80 K in 20 K steps (E∥c). (c) Cooling efficiency $\eta_c(\lambda, T)$ versus excitation wavelength and crystal temperature. The blue and red regions correspond to cooling ($\eta_c$>0) and heating ($\eta_c$<0) regimes, respectively, with the white transition line indicating the local minimum achievable temperature (MAT) at a given wavelength. The global MAT (as indicated by dashed lines) is ~130±10 K at $\lambda$ =2070±0.5 nm which corresponds to the $E_{12} \rightarrow E_{13}$ transition in $Ho^{3+}$ (ref. 20). (d) Ratio of maximum cooling efficiency of Ho:YLF sample over optimal 10% Yb:YLF sample assuming various $\eta_{ext}$ and doping concentrations for Ho:YLF.

**Figure 4: CW-OPO for mid-IR optical refrigeration.** (a) Schematic of CW-OPO design for mid-IR optical refrigeration in Tm and Ho doped crystals. (b) The phase-matching curve of mid-IR CW-OPO. (c) Typical normalized narrow linewidth signal and idler spectra of CW-OPO.

**FIGURES**

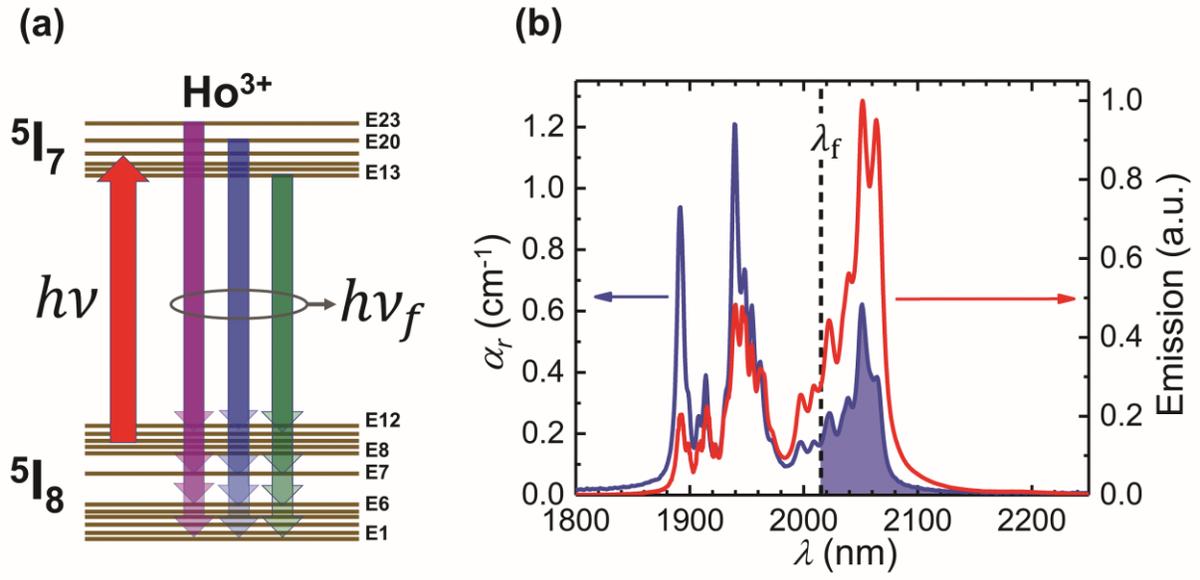

**Figure 1**

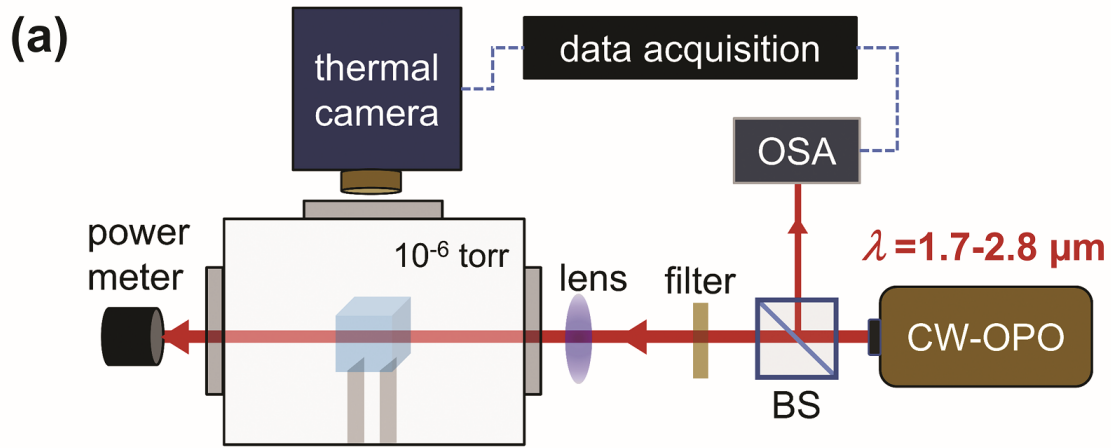

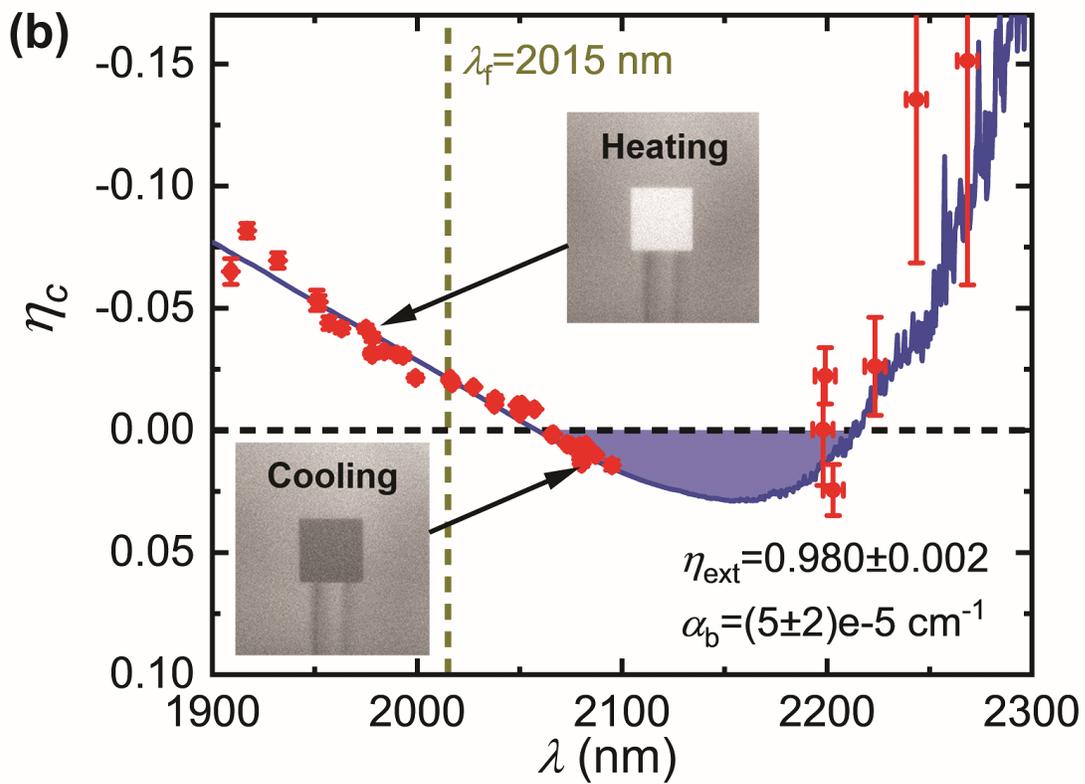

**Figure 2**

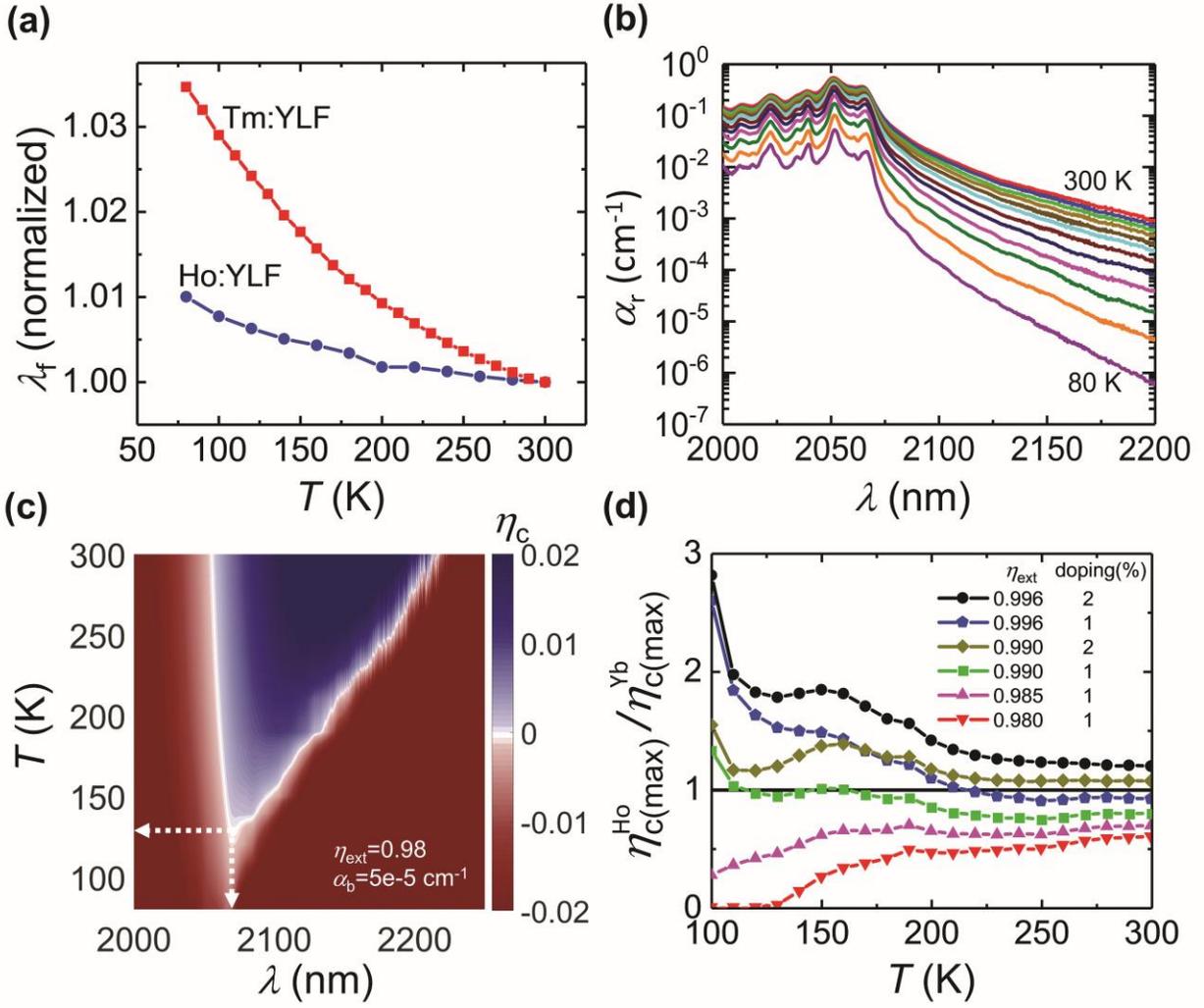

**Figure 3**

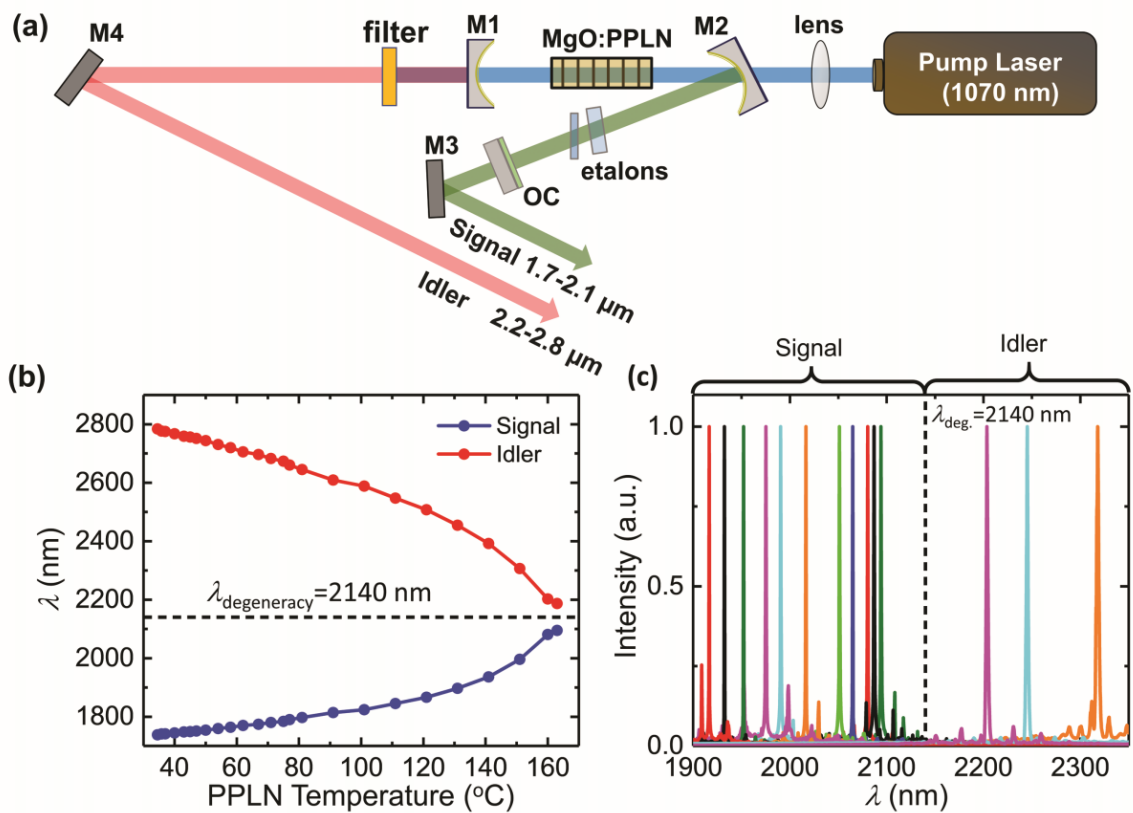

**Figure 4**